\documentclass[preprintnumbers,amsmath,amssymb,12pt,floatfix,epsfig]{revtex4}
\usepackage{dcolumn}
\usepackage{bm}
\usepackage[dvips]{color}
\usepackage{graphicx}
\leftmargin -0.5in \baselineskip=24ptplus.5ptminus.2pt \vspace*{0.5
in} \large
\begin{document}
\title{Extended Optical Model Analyses
of Elastic Scattering and Fusion Cross Section Data for the
$^{7}$Li+$^{208}$Pb System at Near-Coulomb-Barrier Energies using
the Folding Potential} \vspace{0.5cm}

\author{W. Y. So and T. Udagawa}
\address{Department of Physics, University of Texas,
Austin, Texas 78712}

\author{K. S. Kim}
\address{Department of Liberal Arts and Science, Hankuk
Aviation university, Koyang 412-791, Korea}

\author{S. W. Hong and B. T. Kim}
\affiliation{\it Department of Physics and Institute of Basic Science, \\
Sungkyunkwan University, Suwon 440-746, Korea}

\begin{abstract}
\begin{center}
{\vspace{1.0cm} \bf Abstract}
\end{center}
Simultaneous $\chi^{2}$ analyses previously made for elastic
scattering and fusion cross section data for the $^{6}$Li+$^{208}$Pb
system is extended to the $^{7}$Li+$^{208}$Pb system at
near-Coulomb-barrier energies based on the extended optical model
approach, in which the polarization potential is decomposed into
direct reaction (DR) and fusion parts. Use is made of the double
folding potential as a bare potential. It is found that the
experimental elastic scattering and fusion data are well reproduced
without introducing any normalization factor for the double folding
potential and that both the DR and fusion parts of the polarization
potential determined from the $\chi^{2}$ analyses satisfy separately
the dispersion relation. Further, we find that the real part of the
fusion portion of the polarization potential is attractive while
that of the DR part is repulsive except at energies far below the
Coulomb barrier energy. A comparison is made of the present results
with those obtained from the Continuum Discretized Coupled Channel
(CDCC) calculations and a previous study based on the conventional
optical model with a double folding potential. We also compare the
present results for the $^7$Li+$^{208}$Pb system with the analysis
previously made for the $^{6}$Li+$^{208}$Pb system.
\end{abstract}
\maketitle \vspace{1.5cm} PACS numbers : 24.10.-i,~25.70.Jj

\pagebreak

\section{Introduction}

It has been a long standing problem that one is forced to reduce the
strength of the folding potential by a factor $N=0.5 \sim 0.6$ in
order to reproduce the elastic scattering data~\cite{sat1,kee1} for
loosely bound projectiles such as $^{6}$Li and $^{9}$Be within the
optical model approach with a folding potential. This problem has
been ascribed to the strong breakup character of the projectiles;
studies have been made of the effects of the breakup on the elastic
scattering, based on the coupled discretized continuum channel
(CDCC) method~\cite{sak1,kee2}. These studies were very successful
in reproducing the elastic scattering data without introducing any
arbitrary normalization factor and furthermore in understanding the
physical origin of the factor $N=0.5 \sim 0.6$ needed to be
introduced in one-channel optical model calculations. The authors of
Refs.~\cite{sak1,kee2} projected their coupled channel equations to
a single elastic channel equation and deduced the polarization
potential arising from the coupling with the breakup channels. The
resultant real part of the polarization potential was then found to
be repulsive at the surface region around the strong absorption
radius, $R_{sa}$. This shows that the reduction of the folding
potential by a factor of $N=0.5 \sim 0.6$ is to effectively take
into account the effects of this repulsive coupling with the breakup
channels.

In our recent study~\cite{so1}, we explored this problem for the
$^6$Li + $^{208}$Pb system in the framework of the extended
optical model~\cite{uda1,hong,uda2},
in which the optical potential consists of the energy independent
Hartree-Fock part and the energy dependent complex polarization
potential having two components, i.e., the direct reaction (DR)
and fusion parts, which we call the DR and fusion potentials,
respectively. In Ref.~\cite{so1}, using such an extended optical
potential, we performed the simultaneous $\chi^{2}$ analyses of the
elastic scattering and fusion cross section data, determining the
two components of the polarization potentials as functions of the
incident energy. Our expectation was that the resulting real part of
the DR potential would become repulsive consistently with the
results of the CDCC calculations. Indeed the real DR polarization
potential turned out to be repulsive. In addition, it was shown that
both the DR and fusion potentials satisfy the dispersion
relation~\cite{mah1,nag1} separately.

In this work, we extend the analysis made in Ref.~\cite{so1} to the
$^{7}$Li+$^{208}$Pb system. In this system, such a normalization
anomaly as observed in $^{6}$Li+$^{208}$Pb does not appear around
the Coulomb-barrier energies; the normalization factor $N$ necessary
for reproducing the data is close to unity, $N \approx 1$ (see
Ref.~\cite{kee1}), in contrast to the factor $N=0.5 \sim 0.6$ for
$^{6}$Li+$^{208}$Pb.

In Sec. II of this article, we first discuss some characteristic
features of elastic scattering cross section data of
$^{7}$Li+$^{208}$Pb~\cite{kee1} in comparison with those of
$^{6}$Li+$^{208}$Pb. It will be shown in the comparison that the
DR cross section for $^{7}$Li+$^{208}$Pb is expected to be
significantly smaller than that for $^{6}$Li+$^{208}$Pb. In Sec. III, we
first generate the so-called semi-experimental DR cross section,
$\sigma_{D}^{\textrm{semi-exp}}$, from the elastic scattering and
fusion cross section data~\cite{das1, das2}, following the method
described in, e.g., Ref.~\cite{so2}. (Note that use is made of the
fusion cross section data of $^{7}$Li+$^{209}$Bi, since the data are
not available for $^{7}$Li+$^{208}$Pb.) The data of
$\sigma_{D}^{\textrm{semi-exp}}$ are needed for separately
determining the DR and fusion potentials. $\chi^{2}$ analyses of the
elastic scattering, fusion, and semi-experimental DR cross section
data are then carried out in Sec. IV. In Sec. V, a comparison is
made of the present results with those obtained from the CDCC
calculations~\cite{kee2} and a previous study~\cite{kee1} based on
the conventional optical model with a double folding potential. We
also show a comparison of the present result with the analysis
previously made by us~\cite{so1} for the $^{6}$Li+$^{208}$Pb system.
Section VI concludes the paper.

\section{Review of Experimental Cross Sections}

We begin with the discussion of some of the characteristic features
of the elastic scattering cross sections
$d\sigma_{el}/d\sigma_{\Omega}$ data for $^{7}$Li+$^{208}$Pb in
comparison with those for $^{6}$Li+$^{208}$Pb. Such features can
best be seen in the ratio, $P_{E}$, defined by
\begin{equation}
P_{E} \equiv \frac{d\sigma_{el}}{d\sigma_{\Omega}}/
\frac{d\sigma_{C}}{d\sigma_{\Omega}}=d\sigma_{el}/d\sigma_{C}
\end{equation}
as a function of the distance of the closest approach $D$ (or the
reduced distance $d$), where $d\sigma_{C}/d\sigma_{\Omega}$ is the
Coulomb scattering cross section, while $D$ ($d$) is related to the
scattering angle $\theta$ by
\begin{equation}
D=d(A_{1}^{1/3}+A_{2}^{1/3})=\frac{1}{2}D_{0}
\Big[1+\frac{1}{\mbox{sin}(\theta/2)}\Big]
\end{equation}
with
\begin{equation}
D_{0}=\frac{Z_{1}Z_{2}e^{2}} {E},
\end{equation}
$D_{0}$ being the distance of the closest approach in a head-on
collision. Here $(A_{1},Z_{1})$ and $(A_{2},Z_{2})$ are the mass and
charge of the projectile and target ions, respectively, and $E
\equiv E_{c.m.}$ is the incident energy in the center-of-mass
system. $P_{E}$ as defined by Eq. (1) will be referred to as the
elastic probability.

In Figs. 1(a) and 1(b), we present the experimental values of
$P_{E}$ for incident energies around the Coulomb barrier energy as a
function of the reduced distance $d$ for $^{7}$Li+$^{208}$Pb and
$^{6}$Li+$^{208}$Pb, respectively. As seen, the values of $P_{E}$ at
different energies line up to form a very narrow band. This is a
characteristic feature seen in many of the heavy-ion collisions,
reflecting the semiclassical nature of these collisions. $P_{E}$
remains close to unity until two ions approach each other within
a distance $d_{I}$, where $P_{E}$ begins to fall off. The distance
$d_{I}$ is usually called the interaction distance, at which the
nuclear interactions between the colliding ions are switched on, so
to speak. The values of $d_{I}$ are about 1.9~fm for
$^{6}$Li+$^{208}$Pb and 1.8~fm for $^{7}$Li+$^{208}$Pb.

As argued in Ref.~\cite{so2}, the fall off of the $P_{E}$ values in
the region immediately next to $d_{I}$ is due to DR. The fact that the
$d_{I}$-value (1.9~fm) for $^{6}$Li+$^{208}$Pb is larger than the $d_{I}
(1.8$~fm) for $^{7}$Li+$^{208}$Pb shows DR starts to take place at
larger distances for $^{6}$Li+$^{208}$Pb than it does for
$^{7}$Li+$^{208}$Pb. Also, it can be seen that the amount of
decrease of the $P_{E}$ value from unity in $^{6}$Li+$^{208}$Pb is
significantly larger than in $^{7}$Li+$^{208}$Pb at $1.5$~fm $< d <
1.9$~fm, where DR takes place. These features clearly indicate that
DR (which may be dominated by breakup) takes place significantly
stronger in $^{6}$Li+$^{208}$Pb than in $^{7}$Li+$^{208}$Pb. This is
indeed the case as can be seen in the next section from the
semi-experimental DR cross section to be extracted. Finally, we note
that in the region of $d < 1.5$~fm where fusion dominates, the
values of $P_{E}$ for $^{7}$Li+$^{208}$Pb and $^{6}$Li+$^{208}$Pb
are almost identical.

\section{Extracting semi-experimental DR cross section}

For our purpose of determining the fusion and DR potentials
separately, it is desirable to have the data of the DR cross section
in addition to the fusion and elastic scattering cross sections. For
the $^{7}$Li+$^{208}$Pb system, however, no reliable data of the DR
cross sections are available, although considerable efforts have been
devoted to measure the breakup and incomplete fusion cross
sections~\cite{das1,sig1}. Here, we thus generate the so-called
semi-experimental DR cross section $\sigma_{D}^{\textrm{semi-exp}}$,
following the method proposed in Ref.~\cite{so2}.

Our method to generate $\sigma_{D}^{\textrm{semi-exp}}$ resorts to
the well known empirical fact that the total reaction cross section
$\sigma_R$ calculated from the optical model fit to the available elastic
scattering cross section data, $d\sigma_{E}^{\textrm{exp}}/d\Omega$,
usually agrees well with the experimental $\sigma_{R}$, in spite of
the ambiguities in the optical potential. Let us call $\sigma_{R}$
thus generated the semi-experimental reaction cross section
$\sigma_{R}^{\textrm{semi-exp}}$. Then,
$\sigma_{D}^{\textrm{semi-exp}}$ is generated as
\begin{equation}
\sigma_{D}^{\textrm{semi-exp}} = \sigma_{R}^{\textrm{semi-exp}} -
\sigma_{F}^{\textrm{exp}}.
\end{equation}
This approach seems to work even for loosely bound projectiles, as
demonstrated by Kolata {\it et al.}~\cite{kol1} for the
$^{6}$He+$^{209}$Bi system. As already noted in Sec.~I,
$\sigma_F^{exp}$ data are not available $^{7}$Li+$^{208}$Pb, and
thus we use the $\sigma_{F}^{\textrm{exp}}$ data taken for
$^{7}$Li+$^{209}$Bi~\cite{das1, das2}.

Following Ref.~\cite{so2}, we first carry out rather simple optical
model $\chi^{2}$ analyses of elastic scattering data solely for the
purpose of deducing $\sigma_R$ and $\sigma_{R}^{\textrm{semi-exp}}$.
For these preliminary analyses, we assume the optical potential to
be sum of $V_{0}(r)$+$i W_{I}(r)$ and $U_{1}(r,E)$, where $V_{0}(r)$
is the real, energy independent bare folding potential to be
discussed later in Sec. IV. B, $i W_{I}(r)$ is an energy independent
short range imaginary potential to be discussed in Sec. IV. A, and
$U_{1}(r,E)$ is a Woods-Saxon type complex potential with common
geometrical parameters for both real and imaginary parts. The
elastic scattering data are then fitted with a fixed radius
parameter $r_{1}$ for $U_{1}(r,E)$, treating, however, all three
other parameters, the real and the imaginary strengths $V_{1}$ and
$W_{1}$ and the diffuseness parameter $a_{1}$, as adjustable. The
$\chi^{2}$ fitting is done for three choices of the radius
parameter; $r_{1}$=1.3, 1.4, and 1.5 fm. These different choices of
the $r_{1}$-value are made in order to examine the dependence of the
resulting $\sigma_{R}^{\textrm{semi-exp}}$ on the value of $r_1$.

As observed in Ref.~\cite{so2}, the values of
$\sigma_{R}^{\textrm{semi-exp}}$ thus extracted for three different
$r_{1}$-values agree with the average value of
$\sigma_{R}^{\textrm{semi-exp}}$ within 2\%, implying that
$\sigma_{R}^{\textrm{semi-exp}}$ is determined without much
ambiguity. We then identified the average values as the final values
of $\sigma_{R}^{\textrm{semi-exp}}$ at each energy. Using thus
determined $\sigma_{R}^{\textrm{semi-exp}}$, we generated
$\sigma_{D}^{\textrm{semi-exp}}$ by employing Eq.~(4). The resultant
values of $\sigma_{R}^{\textrm{semi-exp}}$ and
$\sigma_{D}^{\textrm{semi-exp}}$ are presented in
Table~\ref{semiexp}, together with $\sigma_{F}^{\textrm{exp}}$. In
Table~\ref{semiexp}, given are also $\sigma_{R}^{\textrm{semi-exp}}$
determined in Ref.~\cite{kee1}. It is noticeable that the two sets of
$\sigma_{R}^{\textrm{semi-exp}}$ determined independently agree
within 1\%. We can also see that the values of
$\sigma_{D}^{\textrm{semi-exp}}$ thus deduced are smaller than
those for $^{6}$Li+$^{208}$Pb~\cite{so1} by a factor of 1.23 $\sim$
1.72 as anticipated from the $P_E$ values discussed in the previous
section.

\begin{table}
\caption{Semi-experimental total reaction and DR cross sections for
the $^{7}$Li+$^{208}$Pb system.} \vspace{2ex} \label{semiexp}
\begin{ruledtabular}
\begin{tabular}{cccccc}
$E_{lab} $ &$E$ & $\sigma_{F}^{\textrm{exp}}$ [11, 12] &
$\sigma_{D}^{\textrm{semi-exp}}$ & $\sigma_{R}^{\textrm{semi-exp}}$
& $\sigma_{R}^{\textrm{semi-exp}}$ [2] \\
(MeV) & (MeV) & (mb) & (mb) & (mb) & (mb) \\ \hline
29 & 28.1 &  18 & 119 & 137 & 138 \\
31 & 30.0 &  88 & 240 & 328 & 327 \\
33 & 31.9 & 218 & 351 & 569 & 572 \\
35 & 33.9 & 366 & 418 & 784 & 787 \\
39 & 37.7 & 650 & 583 &1233 &1242 \\
44 & 42.6 & 866 & 684 &1550 &1553 \\
\end{tabular}
\end{ruledtabular}
\end{table}

\section{Simultaneous $\chi^{2}$ Analyses}

Simultaneous $\chi^{2}-$analyses were then performed on the data
sets of
($d\sigma_{E}^{\textrm{exp}}/d\Omega$,~$\sigma_{D}^{\textrm{semi-exp}}$,
~$\sigma_{F}^{\textrm{exp}}$), by taking the data for
$d\sigma_{E}^{\textrm{exp}}/d\Omega$, and
$\sigma_{F}^{\textrm{exp}}$ from the
literatures~\cite{kee1,das1,das2}. In calculating the $\chi^{2}$
value, we simply assume 1\% errors for all the experimental data.
The 1\% error is about the average of errors in the measured elastic
scattering cross sections, but much smaller than the errors in the
DR ($\sim$5\%) and fusion ($\sim$10\%) cross sections. Assigning the
1\% error for DR and fusion cross sections is thus equivalent to
increasing the weight for the DR and fusion cross sections in
evaluating the $\chi^{2}$-values by factors of 25 and 100,
respectively. Such a choice of errors may be reasonable, since we
have only one datum point for each of these cross sections, while
there are more than 50 data points for the elastic scattering cross
sections.

\subsection{Necessary Formulae}

The optical potential $U(r,E)$ we use in the present work has the
following form;
\begin{equation} \label{e-full_poten}
U(r;E) = V_{C}(r)-[V_{0}(r)+U_{F}(r;E)+U_{D}(r;E)],
\end{equation}
where $V_{C}(r)$ is the usual Coulomb potential with $r_{C}$=1.25 fm
and $V_{0}(r)$ is the bare nuclear potential, for which use is made
of the double folding potential to be described in the next
subsection. $U_{F}(r;E)$ and $U_{D}(r;E)$ are, respectively, fusion
and DR parts of the so-called polarization potential~\cite{love}
that originates from couplings to the respective reaction channels.
Both $U_{F}(r;E)$ and $U_{D}(r;E)$ are complex and their forms are
assumed to be of volume-type and
surface-derivative-type~\cite{hong,kim1}, respectively. $U_{F}(r;E)$
and $U_{D}(r;E)$ are explicitly given by
\begin{equation} \label{e-u_f}
U_{F}(r;E) = [V_{F}(E)+iW_{F}(E)]f(X_{F})+iW_{I}(r),
\end{equation}
and
\begin{equation}
U_{D}(r;E) = [V_{D}(E)+iW_{D}(E)]4a_{D}\frac{df(X_{D})}{dR_{D}},
\vspace{2ex}
\end{equation}
where $f(X_{i})=[1+\mbox{exp}(X_{i})]^{-1}$ with
$X_{i}=(r-R_{i})/a_{i}$ $({\it i}=F\; \mbox{and} \; D)$ is the usual
Woods-Saxon function with the fixed geometrical parameters of
$r_{F}=1.40$~fm, $a_{F}=0.33$~fm, $r_{D}=1.47$~fm, and
$a_{D}=0.56$~fm, while $V_{F}(E)$, $V_{D}(E)$, $W_{F}(E)$, and
$W_{D}(E)$ are the energy-dependent strength parameters. Since we
assume the geometrical parameters to be the same for both the real
and imaginary potentials, the strength parameters $V_{i}(E)$ and
$W_{i}(E)$ ($i=F$ or $D$) are related through a dispersion
relation~\cite{mah1},
\begin {equation} \label{e-disper}
V_{i}(E)=V_{i}(E_{s}) + \frac {E-E_{s}}{\pi } \mbox{P}
\int_{0}^{\infty} dE' \frac {W_{i}(E')}{(E'-E_{s})(E'-E)},
\vspace{2ex}
\end {equation}
where P stands for the principal value and $V_{i}(E_{s})$ is the
value of $V_{i}(E)$ at a reference energy $E=E_{s}$. Later, we will
use Eq.~(\ref{e-disper}) to generate the final real strength
parameters $V_{F}(E)$ and $V_{D}(E)$ using $W_{F}(E)$ and $W_{D}(E)$
fixed from the $\chi^{2}$ analyses. Note that the breakup cross
section may include contributions from both Coulomb and nuclear
interactions, which implies that the direct reaction potential
includes effects coming from not only the nuclear interaction, but
also from the Coulomb interaction.

The last imaginary potential $W_{I}(r)$ in $U_{F}(r;E)$ given by
Eq.~(\ref{e-u_f}) is a short-range potential of the Woods-Saxon type
given as
\begin{equation}
W_{I}(r) = W_{I}f(X_{I}),
\end{equation}
with $W_{I}=40$~MeV, $r_{I}=0.8$~fm, and $a_{I}=0.30$~fm. This
imaginary potential was first introduced~\cite{so1} in order to
eliminate unphysical reflection in the radial wave functions of low
partial waves when this $W_{I}(r)$ is absent. Because of the large
strength of the folding potential $V_{0}$ used in this study and
also because $W_{F}(E)f(X_{F})$ of Eq.~(\ref{e-u_f}) turns out to be
not so strong enough, reflections of lower partial waves appear in the
asymptotic region, which causes unphysical oscillations of
differential elastic cross sections at large angles, particularly at
relatively high energies above the Coulomb-barrier, but physically
such reflection should not occur because of the strong absorption
that should exist inside the nucleus. $W_{I}(r)$ is thus introduced
in order to take care of the strong absorption inside and eliminate
this unphysical effect. We might then need to introduce a real part
$V_{I}(r)$ corresponding to $W_I (r)$, but we ignored the real part,
simply because such a real potential did not affect at all the real
physical observables, which means that it is impossible to extract
the information on $V_{I}(r)$ from the analyses of the experimental
data. Further, as will be discussed later in Sec. IV E, $W_{I}(r)$
is also insensitive to the observables, particularly at low energies
around and below the Coulomb-barrier. This means that it is also
impossible to extract information of the energy dependence of
$W_{I}(r)$ from the data. For this reason, we simply ignore in this
study the energy dependence of $W_{I}(r)$.

In the extended optical model, fusion and DR cross sections,
$\sigma_{F}$ and $\sigma_{D}$, respectively, are calculated by using
the following expression~\cite{uda1,hong,uda2,huss}
\begin {equation}
\sigma_{i} = \frac {2}{\hbar v} <\chi^{(+)}|
\mbox{Im}~[U_{i}(r;E)]|~\chi^{(+)}> \hspace{.5in}
(i=F\;\mbox{or}\;D),
\end{equation}
where $\chi^{(+)}$ is the usual distorted wave function that
satisfies the Schr\"{o}dinger equation with the full optical model
potential $U(r;E)$ in Eq.~(\ref{e-full_poten}). $\sigma_{F}$ and
$\sigma_{D}$ are thus calculated within the same framework as
$d\sigma_{el}/d\Omega$ is calculated. Such a unified description
enables us to evaluate all the different types of cross sections on
the same footing.

\subsection{The Folding Potential}

The double folding potential $V_{0}(r)$ we use in the present study
as the bare potential may be written as~\cite{sat1}
\begin{equation}
V_{0}(r)=\int d{\bf r}_{1} \int d{\bf r}_{2} \rho_{1}(r_{1})
\rho_{2}(r_{2}) v_{NN}(r_{12}=|\bf{r}-\bf{r}_{1}+\bf{r}_{2}|),
\end{equation}
where $\rho_{1}(r_{1})$ and $\rho_{2}(r_{2})$ are the nuclear matter
distributions for the target and projectile nuclei, respectively,
while $v_{NN}$ is the sum of the M3Y interaction that describes the
effective nucleon-nucleon interaction and the knockon exchange
effect given as
\begin{equation}
v_{NN}(r)=7999\frac{e^{-4r}}{4r}-2134\frac{e^{-2.5r}}{2.5r}-262
\delta (r).
\end{equation}
We use for $\rho_{1}(r)$ the following Woods-Saxon form taken from
Ref.~\cite{jag1}
\begin{equation}
\rho_{1}(r)=\rho_{0}/\left[1+\mbox{exp}\left(\frac{r-c}{z}\right)\right],
\end{equation}
with $c=6.624$~fm and $z=0.549$~fm, while for $\rho_{2}(r)$ the
following is taken from Ref.~\cite{coo1};
\begin{equation}
\rho_{2}(r)=(A+Br^{2}) e^{-\alpha^{2}r^{2}},
\end{equation}
with $A$=0.13865~fm$^{-3}$, $B$=0.02316~fm$^{-1}$, and
$\alpha$=0.578~fm$^{-1}$. We then use the code DFPOT of
Cook~\cite{coo2} for evaluating $V_{0} (r)$.

\subsection{Threshold Energies of Subbarrier Fusion and DR}

As in Ref.~\cite{so1}, we utilize as an important ingredient the
so-called threshold energies $E_{0,F}$ and $E_{0,D}$ of subbarrier
fusion and DR, respectively, which are defined as zero intercepts of
the linear representation of the quantities $S_{i}(E)$, defined by
\begin{equation} \label{e-s_factor}
S_{i} \equiv \sqrt{E \sigma_{i}} \approx \alpha_{i} (E-E_{0,i})
\;\;\; (i=F \; \mbox{or} \; D),
\end{equation}
where $\alpha_{i}$ is a constant. $S_{i}$ with $i=F$, i.e., $S_{F}$
is the quantity introduced originally by Stelson {\it et
al.}~\cite{stel}, who showed that in the subbarrier region $S_{F}$
from the measured $\sigma_{F}$ can be represented very well by a
linear function of $E$ (linear systematics) as in
Eq.~(\ref{e-s_factor}). In Ref.~\cite{kim1}, we extended the linear
systematics to DR cross sections. In fact the DR data are also well
represented by a linear function.

In Fig.~\ref{s-factor}, we present the experimental $S_{F}(E)$ and
$S_{D}(E)$. For $S_{D}(E)$, use is made of
$\sigma_{D}^{\textrm{semi-exp}}$. From the zeros of $S_{i}(E)$, one
can deduce $E_{0,D}^{\textrm{semi-exp}}$=19.3~MeV and
$E_{0,F}^{\textrm{exp}}=$26.5~MeV. For both $i=F$ and $D$, the
observed $S_{i}$ are very well approximated by straight lines in the
subbarrier region and thus $E_{0,i}$ can be extracted without much
ambiguity. It is worthwhile to remark that
$E_{0,D}^{\textrm{semi-exp}}$ is found to be considerably smaller
than $E_{0,F}^{\textrm{exp}}$, implying that DR channels open at
smaller energies than fusion channels, which seems physically
reasonable.

$E_{0,i}$ may then be used as the energy where the imaginary
potential $W_{i}(E)$ becomes zero, i.e.,
$W_{i}(E_{0,i})=0$~\cite{kim1,kim2}. This procedure will be used
later in the next subsection for obtaining a mathematical expression
for $W_{i}(E)$.

\subsection{$\chi^{2}$ Analyses}

All the $\chi^{2}$ analyses performed in the present work are
carried out by using the folding potential as its bare potential
$V_{0}(r)$ described in Sec. III. B and by using the fixed
geometrical parameters for the polarization potentials,
$r_{F}$=1.40~fm, $a_{F}$=0.33~fm, $r_{D}$=1.47~fm, and
$a_{D}$=0.56~fm, which are close to the values used in our previous
study~\cite{kim1}.
A slight change of the values used in Ref.~\cite{kim1} is made in
order to improve the $\chi^{2}$ fitting.

As in Ref.~\cite{kim1}, the $\chi^{2}$ analyses are done in two
steps; in the first step, all 4 strength parameters, $V_{F}(E)$,
$W_{F}(E)$, $V_{D}(E)$ and $W_{D}(E)$ are varied. In this step, we
could fix fairly well the strength parameters of the DR potential,
$V_{D}(E)$ and $W_{D}(E)$, in the sense that $V_{D}(E)$ and
$W_{D}(E)$ were determined as a smooth function of $E$. The values
of $V_{D}(E)$ and $W_{D}(E)$ thus extracted are presented in
Fig.~\ref{dispersion} by open circles. The values of $W_{D}(E)$
can be well represented by the following function of $E$
(in units of MeV)
\begin{equation} \label{e-W_D}
W_{D}(E) \; = \; \left \{
\begin{array}{lll}
0 &\;\; \mbox{for $E\leq E_{0, D}^{\textrm{semi-exp}}=$19.3} \\
0.075(E-19.3) &\;\; \mbox{for 19.3$<E\leq$29.3} \\
0.75 &\;\; \mbox{for 29.3$< E$} \\
\end{array}
\right. \vspace{2ex}
\end{equation}
Note that the threshold energy where $W_{D}(E)$ becomes zero is set
equal to $E_{0,D}^{\textrm{semi-exp}}$ as determined in the previous
subsection and is indicated by the open circle at $E=19.3$~MeV in
Fig.~\ref{dispersion}. The dotted line in the lower panel of
Fig.~\ref{dispersion} represents Eq.~(\ref{e-W_D}), while that in
the upper panel of Fig.~\ref{dispersion} denotes $V_{D}$ as
calculated by the dispersion relation Eq.~(\ref{e-disper}), with
$W_{D}(E)$ given by Eq.~(\ref{e-W_D}). As seen, the dotted lines
reproduce the open circles quite well, indicating that $V_{D}(E)$
and $W_{D}(E)$ extracted by the $\chi^{2}$ analyses satisfy the
dispersion relation.

In this first step of $\chi^{2}$ fitting, however, the values of
$V_{F}(E)$ and $W_{F}(E)$ are not reliably fixed in the sense that
the extracted values fluctuate considerably as functions of $E$.
This is understandable from the expectation that the elastic
scattering data can probe most accurately the optical potential in
the peripheral region, which is nothing but the region characterized
by the DR potential. The part of the nuclear potential responsible
for fusion is thus difficult to pin down in this first step.

In order to obtain more reliable information on $V_{F}$ and $W_{F}$,
we thus performed the second step of the $\chi^{2}$ analysis; this
time, instead of doing a 4-parameter search we fixed $V_{D}$ and
$W_{D}$ as determined by the first $\chi^{2}$ fitting, i.e.,
$W_{D}(E)$ given by Eq.~(\ref{e-W_D}) and $V_{D}(E)$ predicted from
the dispersion relation. We then performed 2-parameter $\chi^{2}$
analyses, treating only $V_{F}(E)$ and $W_{F}(E)$ as adjustable
parameters. The values thus determined are presented in
Fig.~\ref{dispersion} by filled circles. As seen, both $V_{F}(E)$ and
$W_{F}(E)$ are determined to be fairly smooth functions of $E$. The
$W_{F}(E)$ values may be represented by
\begin{equation} \label{e-W_F}
W_{F}(E) \; = \; \left \{ \begin{array}{lll}
0 &\;\; \mbox{for $E\leq E_{0, F}^{\textrm{exp}}=$26.5} \\
0.588(E-26.5) &\;\; \mbox{for 26.5$<E\leq$29.9} \\
2.00 &\;\; \mbox{for 29.9$< E$} \\
\end{array}
\right. \vspace{2ex}
\end{equation}
As is done for $W_{D}(E)$, the threshold energy where $W_{F}(E)$
becomes zero is set equal to $E_{0,F}^{\textrm{exp}}$ which is also
indicated by the filled circle in Fig.~\ref{dispersion}. As seen, the
$W_{F}(E)$ values determined by the second $\chi^{2}$ analyses can
fairly well be represented by the functions given by
Eq.~(\ref{e-W_F}). Note that the energy variations seen in
$W_{F}(E)$ and $V_{F}(E)$ are more rapid compared to those seen in
$W_{D}(E)$ and $V_{D}(E)$, and are similar to those observed with
tightly bound projectiles~\cite{bae1,lil1,ful1}. It is thus seen
that the resultant $V_{F}(E)$ and $W_{F}(E)$ exhibit the threshold
anomaly.

Using $W_{F}(E)$ given by Eq.~(\ref{e-W_F}), one can generate
$V_{F}(E)$ from the dispersion relation. The results are shown by
the solid curve in the upper panel of Fig.~\ref{dispersion}, which
again well reproduces the values extracted from the $\chi^{2}$
fitting. This means that the fusion potential determined from the
present analysis also satisfies the dispersion relation.

\subsection{Final Calculated Cross Sections in Comparison with
the Data}

Using $W_{D}(E)$ given by Eq.~(\ref{e-W_D}) and $W_{F}(E)$ given by
Eq.~(\ref{e-W_F}) together with $V_{D}(E)$ and $V_{F}(E)$ generated
by the dispersion relation, we have performed the final calculations
of the elastic, DR and fusion cross sections. The results are
presented in Figs.~\ref{elastic} and~\ref{reaction} in comparison
with the experimental data. All the data are well reproduced by the
calculations.

It may be worth noting here that the theoretical fusion cross
section, $\sigma_{F}^{\textrm{th}}$, includes contributions from two
imaginary components $W_{I} (r)$ and $W_{F} (E) f(X_{F})$ in
$U_{F}(r,E)$ of Eq.~(\ref{e-u_f}). In Table~\ref{fusion} the partial
contributions from the $W_{I}(r)$ part, denoted by $\sigma_{I}$, are
presented in comparison with the total calculated fusion cross
section, $\sigma_{F}^{\textrm{th}}$. As seen, the contribution from
the inner part, $W_{I}$, amounts to $22 \sim 46$~\% of
$\sigma_{F}^{\textrm{th}}$, which is relatively small but not
negligible at all.

In spite of this non-negligible contribution from $W_{I}(r)$,
$W_{I}(r)$ is rather insensitive to the final
total fusion cross section, $\sigma^{\textrm{th}}_{F}$, and also to
the elastic scattering cross sections, particularly in the energy
region where the strength of $W_{F}(E)$ varies rapidly with $E$. To
see this, we have repeated the cross section calculations by
reducing the value of $W_{I}$ to 20~MeV at
$E=28.1$~MeV. This energy is the lowest energy considered in the
present study and is a typical energy in the region
where $W_{F}(E)$ changes rapidly with $E$. The
resulting elastic scattering cross section is found to remain
essentially the same. The value of $\sigma_{I}$ decreases from 11 mb
to 10 mb, and $\sigma_{F}$ increases from 13 mb to 14 mb,
leaving the total fusion cross section,
$\sigma^{\textrm{th}}_{F}$, unchanged. This result
confirms what was stated earlier in Sec. IV A that it is impossible
to extract information of the energy dependence of $W_{I}$ from the
analysis of the experimental data, justifying the present approach
to treat $W_{I}$ as a constant.

\begin{table}
\caption{Partial contributions $\sigma_{I}$ and $\sigma_{F}$ to the
fusion cross sections.} \vspace{2ex} \label{fusion}
\begin{ruledtabular}
\begin{tabular}{ccccc}
$E_{lab} $ &$E$ & $\sigma_{I}$ & $\sigma_{F}$ & $\sigma_{F}^{\textrm{th}}$ \\
(MeV) & (MeV) & (mb) & (mb) & (mb) \\ \hline
29 & 28.1 & 11 & 13 & 24 \\
31 & 30.0 & 23 & 80 & 103 \\
33 & 31.9 & 53 & 166 & 219 \\
35 & 33.9 & 91 & 259 & 350 \\
39 & 37.7 & 175 & 430 & 605 \\
44 & 42.6 & 277 & 604 & 881 \\
\end{tabular}
\end{ruledtabular}
\end{table}

\subsection{Discussions}
As already remarked in Sec. IV. D, the real and imaginary parts of
both DR and fusion polarization potentials determined from the
present $\chi^{2}$ analyses satisfy the dispersion
relation~\cite{mah1,nag1} separately. Furthermore, the fusion
potential exhibits the threshold anomaly as observed in heavy ion
collisions involving tightly bound
projectiles~\cite{bae1,lil1,ful1}.
For the $^{6}$Li+$^{208}$Pb system studied earlier~\cite{so1}
similar threshold anomaly for the fusion potential and the
dispersion relation were observed.

It is remarkable that the real part of the DR potential, which we
denote here by $V_{D}(r,E)$, turns out to be repulsive at most of
the energies considered; only exceptions appear at the lowest energy
point of $E$=28.1~MeV, where $V_{D}(r,E)$ becomes very weakly
attractive (see Fig.~3). The final dispersive $V_{D}(r,E)$
determined by using the dispersion relation, Eq.~(\ref{e-disper}),
with $W_{D}(E)$ given by Eq.~(\ref{e-W_D}) is repulsive above $E
\simeq$ 29 MeV, but becomes attractive between $E=$ 19 MeV and 29
MeV. We remark that the repulsiveness of $V_{D}(r,E)$ for
$^{7}$Li+$^{208}$Pb is considerably weaker than that for
$^{6}$Li+$^{208}$Pb~\cite{so1}, 
which is consistent with the results drawn from
the CDCC study made in Ref.~\cite{kee2}, where the polarization
potentials due to the coupling to the breakup channels are
calculated for both $^{6}$Li+$^{208}$Pb and $^{7}$Li+$^{208}$Pb.

It is also remarkable that the polarization potential
in the surface region, say at the strong absorption radius of
$R_{sa}=12.4$~fm, are dominated by the DR part of the
potential as shown in Fig.~\ref{imaginary}. (Note that
Fig.~\ref{dispersion} shows only the potential strength parameters,
not the potential values.) The same was true for $^{6}$Li+$^{208}$Pb
in Ref.~\cite{so1}. Let us take as an example the imaginary part of
the potential. Then the contribution to the total imaginary part of
the potential from the fusion part is less than 6\% and 15\% for
$^{7}$Li+$^{208}$Pb and $^{6}$Li+$^{208}$Pb systems, respectively.
Therefore, the total polarization potential in the surface region is
mainly characterized by the DR potential.

It is then interesting to compare the values of the total imaginary
potential at $r=R_{sa}$, $W(r=R_{sa},E)$, with those obtained in
Ref.~\cite{kee1}, where the $\chi^{2}$ analyses of the elastic
scattering data of both $^{6}$Li+$^{208}$Pb and $^{7}$Li+$^{208}$Pb
were carried out by using double folding potentials as a real
potential and a Woods-Saxon type as an imaginary potential. In
Ref.~\cite{kee1} the overall normalization constant $N$ of the
double folding potential and all three parameters (the strength,
radius, and diffuseness parameters) of the imaginary potential were
treated as adjustable parameters. An important conclusion drawn from
the analyses was that the resultant potentials at the surface
exhibit the threshold anomaly for $^{7}$Li but not for $^{6}$Li.

In Fig.~\ref{potential}, presented are values of $W(r,E)$, at
$r=R_{sa}=12.4$~fm obtained directly from the $\chi^{2}$ analyses
(not those of the dispersive potential such as given by
Eqs.~(\ref{e-W_D}) and (\ref{e-W_F})) carried out here for $^{7}$Li
and in Ref.~\cite{so1} for $^{6}$Li in comparison with those taken
from Fig.~2 of Ref.~\cite{kee1}. Note that the potential values
taken from Ref.~\cite{kee1} are multiplied by factors 1.23 and 1.11
for $^{7}$Li and $^{6}$Li, respectively, for comparison.
Figure~\ref{potential} shows that the two sets of the values are
very close to each other, demonstrating clearly that the energy
dependence of the $W(R_{sa},E)$ values determined in both cases are
essentially the same. Combined with the above mentioned fact that
the $W(R_{sa},E)$ values determined in the present study and in
Ref.~\cite{so1} are essentially those of the DR potential, it
follows that the energy dependence seen in the $W(R_{sa},E)$ values
of Ref.~\cite{kee1} is that of DR. In this sense, the threshold
anomaly claimed to be seen in Ref.~\cite{kee1} for $^{7}$Li is not
the threshold anomaly due to fusion that copiously observed in the
tightly bound projectiles~\cite{bae1,lil1,ful1}.

\section{Conclusions}

Simultaneous $\chi^{2}$ analyses are made for elastic scattering and
fusion cross section data for the $^{7}$Li+$^{208}$Pb system at
near-Coulomb-barrier energies based on the extended optical model
approach in which the polarization potential is decomposed into
DR and fusion parts. Use is made of the double
folding potential as a bare potential. It is found that the
experimental elastic scattering and fusion data are well reproduced
without introducing any normalization factor for the double folding
potential and also that both DR and fusion parts of the polarization
potential determined from the $\chi^{2}$ analyses satisfy separately
the dispersion relation. Moreover, we find that the real part of the
fusion portion of the polarization potential is attractive while
that of the DR part is repulsive except at energies far below the
Coulomb barrier energy. The repulsive real part of the DR potential
is, however, considerably smaller than that for $^{6}$Li+$^{208}$Pb
obtained earlier~\cite{so1}, reflecting the fact that the DR
(breakup) cross section for $^{7}$Li+$^{208}$Pb is smaller than that
for $^{6}$Li+$^{208}$Pb. Accordingly, the imaginary part of the DR
potential obtained for $^{7}$Li+$^{208}$Pb is smaller than that for
$^{6}$Li+$^{208}$Pb. 
These features of the polarization potential remarked above 
are qualitatively consistent with those obtained in the CDCC 
calculation~\cite{kee1}.

We find that the energy dependence of the optical potential
determined in Ref.~\cite{kee1} is very much like that of the DR
potential deduced in the present study. This means that the energy
dependence seen in Ref.~\cite{kee1} is not a real threshold anomaly
due to fusion, but much slowly varying energy dependence due to DR.

\acknowledgments This work was supported in part by the Korea
Research Foundation Grants funded by the Korean Governement (MOEHRD)
(KRF-2006-214-C00014 and KRF-2003-070-C00015). It was also supported
in part by the Korea Science and Engineering Foundation grant funded
by the Korea Government (MOST) (No. M20608520001-06B0852-00110).

\newpage

\begin{figure}
\begin{center}
\includegraphics[width=0.85\linewidth] {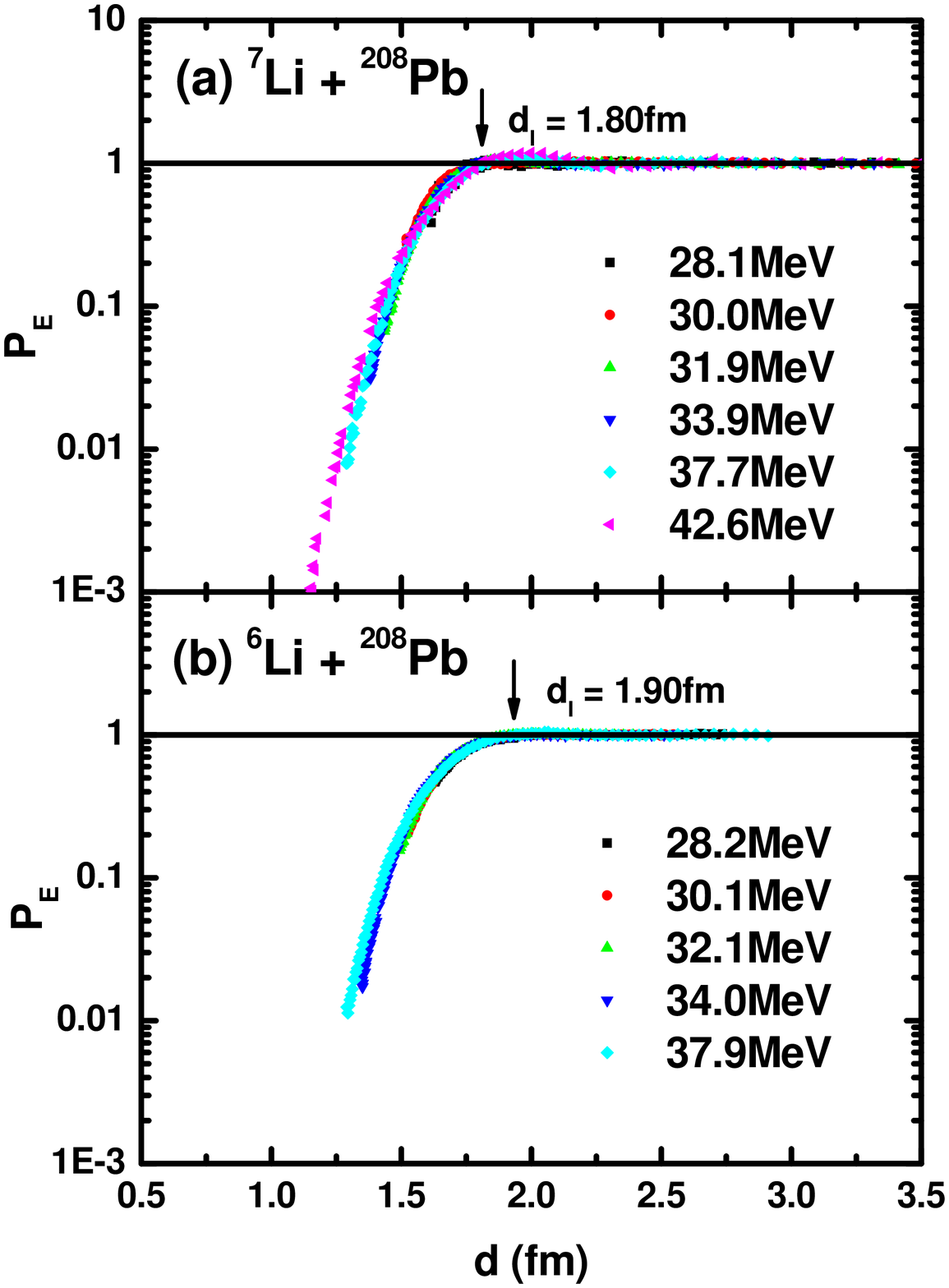}
\end{center}
\caption{\label{pe_value} (Color online) $P_{E}$ values for (a) the
$^{7}$Li+$^{208}$Pb system and (b) the $^{6}$Li+$^{208}$Pb system.}
\end{figure}

\begin{figure}
\begin{center}
\vspace{5.0cm}
\includegraphics[width=0.95\linewidth] {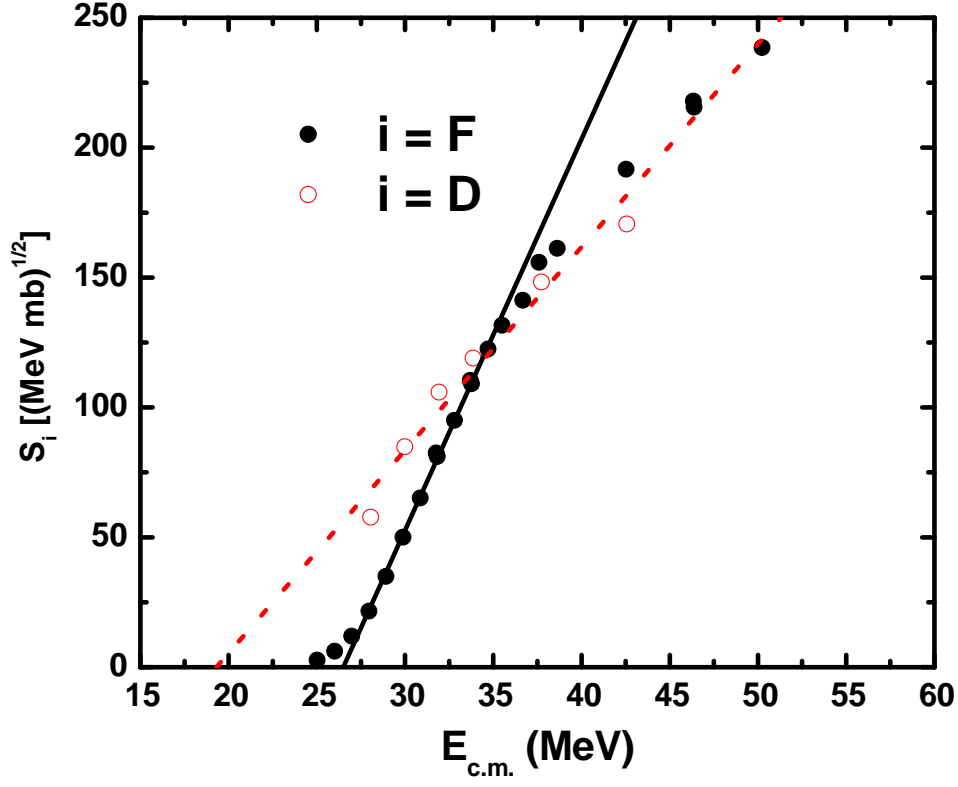}
\end{center}
\caption{\label{s-factor} (Color online) The Stelson plot of
$S_{i}=\sqrt{E~ \sigma_{i}}$ for DR ($i=D$, open circles) and fusion
($i=F$, filled circles) cross sections. Use is made of the
semi-experimental DR cross section for $S_{D}$, while the
experimental fusion cross section is employed for $S_{F}$. 
The intercepts of the straight lines with the energy axis give us 
the threshold energies $E_{0,D}^{\textrm{semi-exp}}$ = 19.3 MeV 
and $E_{0,F}^{\textrm{exp}}$ = 26.5 MeV.}
\end{figure}

\begin{figure}
\begin{center}
\includegraphics[width=0.70\linewidth]{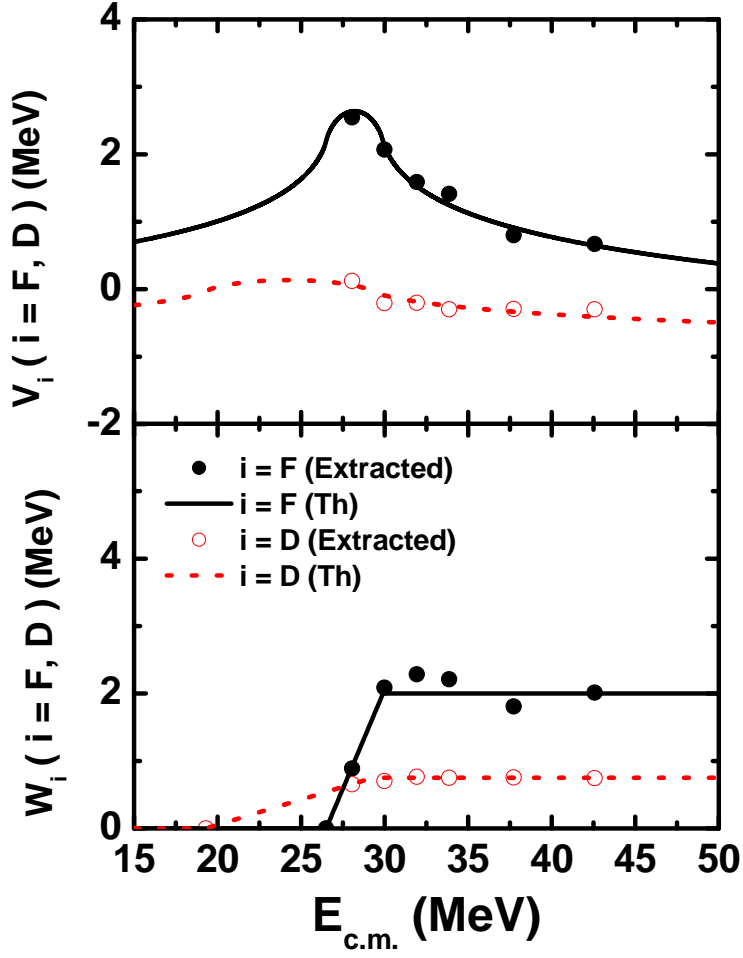}
\end{center}
\caption{\label{dispersion}~(Color online) The strength parameters
$V_{i}$ (upper panel) and $W_{i}$ (lower panel) for $i=D$ and $F$ as
functions of $E_{c.m.}$. The open and filled circles are the strength
parameters for $i=D$ and $F$, respectively. The dotted and solid
lines in the lower panel denote $W_{D}$ and $W_{F}$ from Eqs.
(\ref{e-W_D}) and (\ref{e-W_F}), respectively, while the dotted and
solid curves in the upper panel represent $V_{D}$ and $V_{F}$
calculated by using the dispersion relation of Eq. (\ref{e-disper})
with $W_{i}$ given by Eqs. (\ref{e-W_D}) and (\ref{e-W_F}). The
potential values and the corresponding reference energies 
used in Eq.~(8) are such that $V_F$ ($E_s$=29.9MeV) = $2.2$ MeV 
and $V_D$ ($E_s$=29.3MeV) = $-0.03$ MeV, respectively.}
\end{figure}

\begin{figure}
\begin{center}
\includegraphics[width=0.85\linewidth]{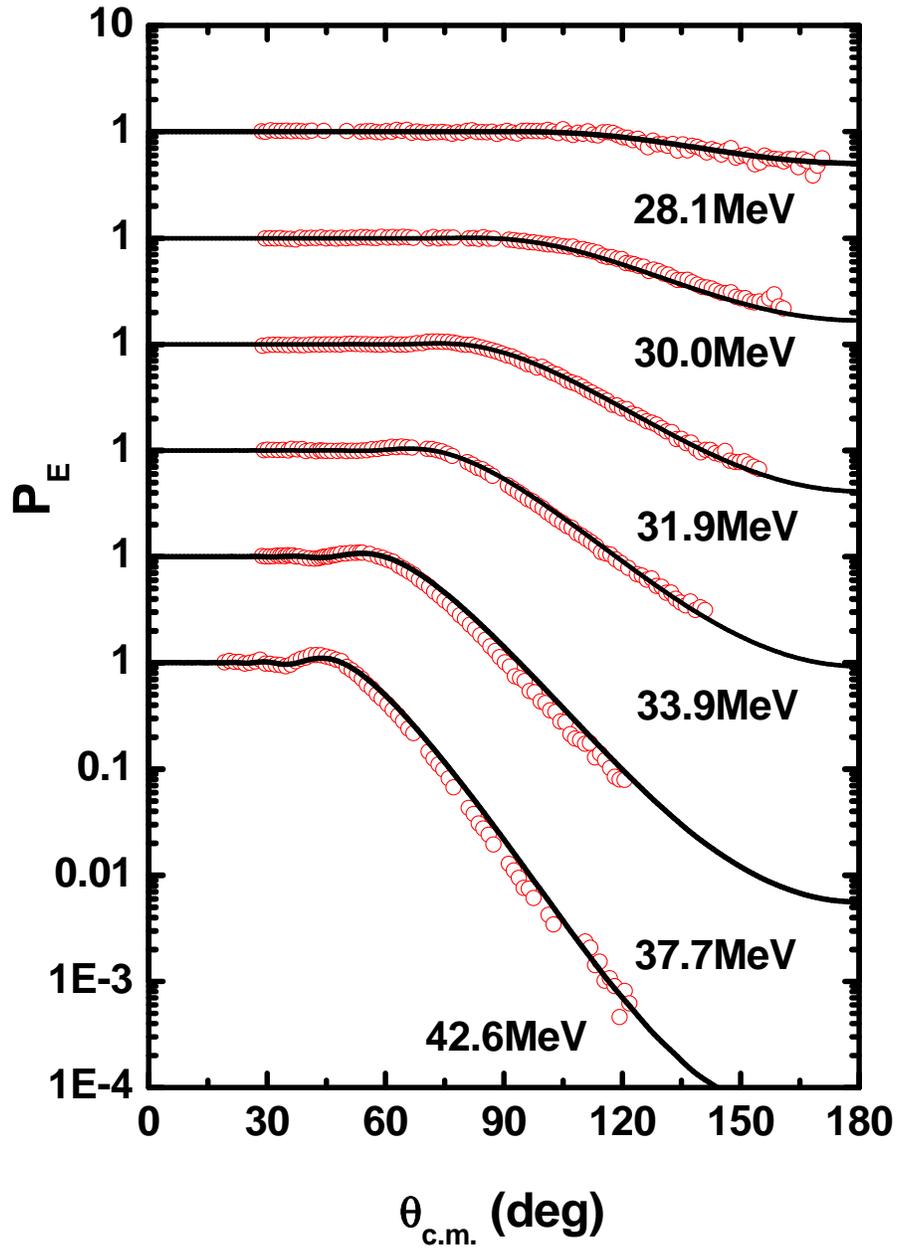}
\end{center}
\caption{\label{elastic}~(Color online) Ratios of the elastic
scattering cross sections to the Rutherford cross section calculated
with our final dispersive optical potential are shown in comparison
with the experimental data. The data are taken from
Ref.~\cite{kee1}.}
\end{figure}

\begin{figure}
\begin{center}
\vspace{5.0cm}
\includegraphics[width=0.95\linewidth]{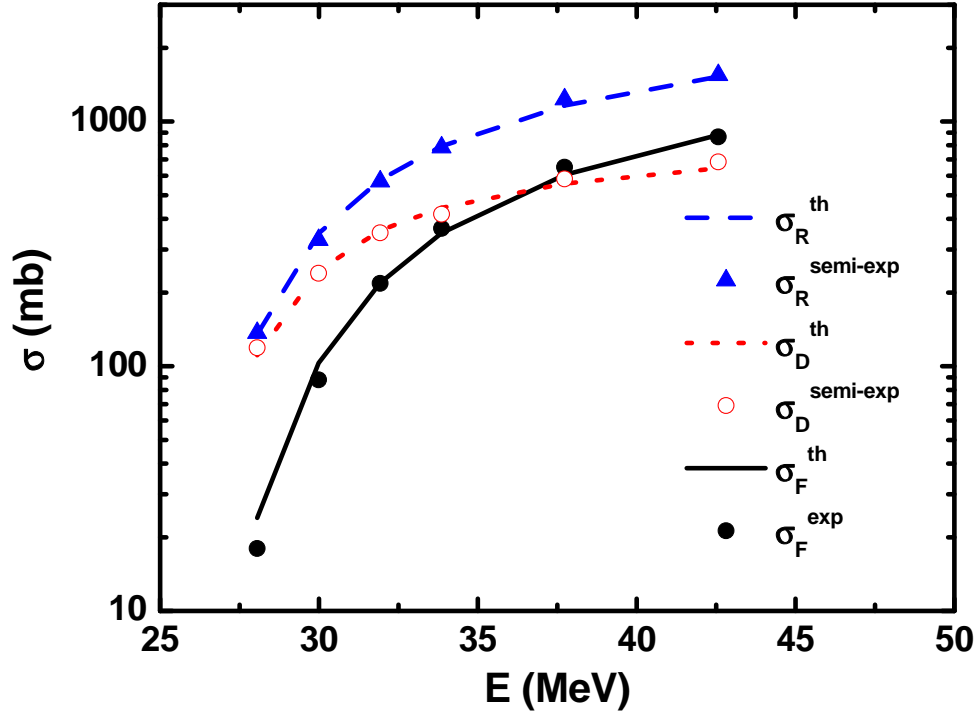}
\end{center}
\caption{\label{reaction}~(Color online) DR and fusion cross
sections calculated with our final dispersive optical potentials are
shown in comparison with the experimental data.
$\sigma_{D}^{\textrm{semi-exp}}$ denoted by the open circles are
obtained as described in Sec.II. The fusion data are from
Refs.~\cite{das1,das2}.}
\end{figure}

\begin{figure}
\begin{center}
\vspace*{5.0cm}
\includegraphics[width=0.95\linewidth]{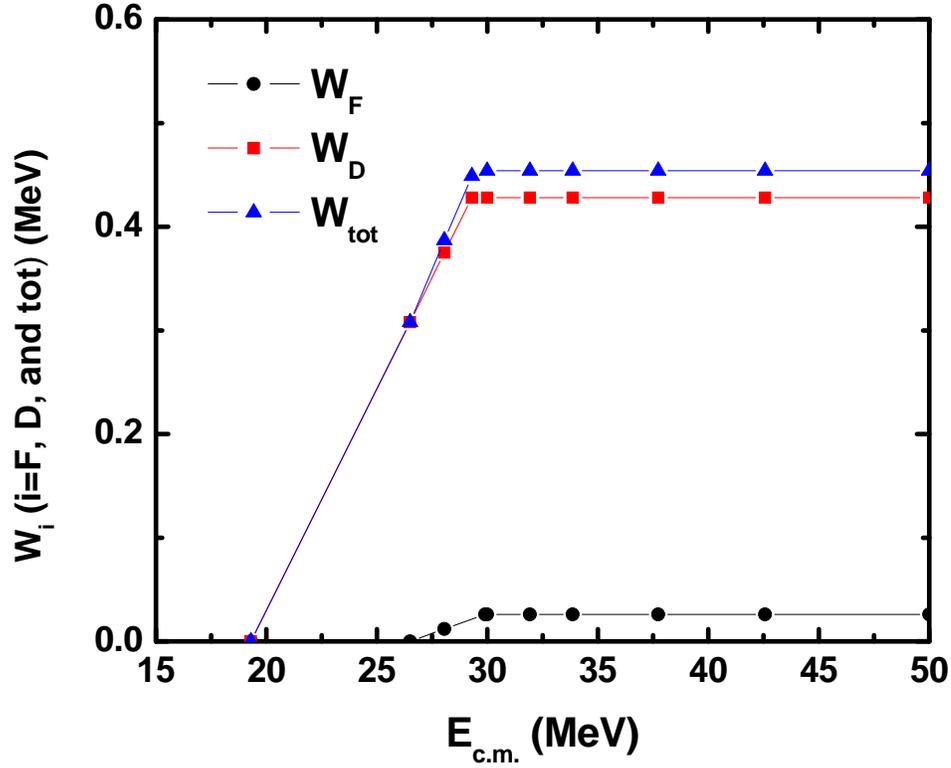}
\end{center}
\caption{\label{imaginary} (Color online) The values of $W_{F}(r,
E)$, $W_{D}(r, E)$ and the sum $W_{tot}(r, E)=W_{F}(r, E)+W_{D}(r,
E)$ as functions of $E$ calculated by using Eqs.~(6), (7), (16) and (17) at the strong absorption radius, $r = R_{sa}$=12.4fm, for all energies.}
\end{figure}

\begin{figure}
\begin{center}
\includegraphics[width=0.85\linewidth]{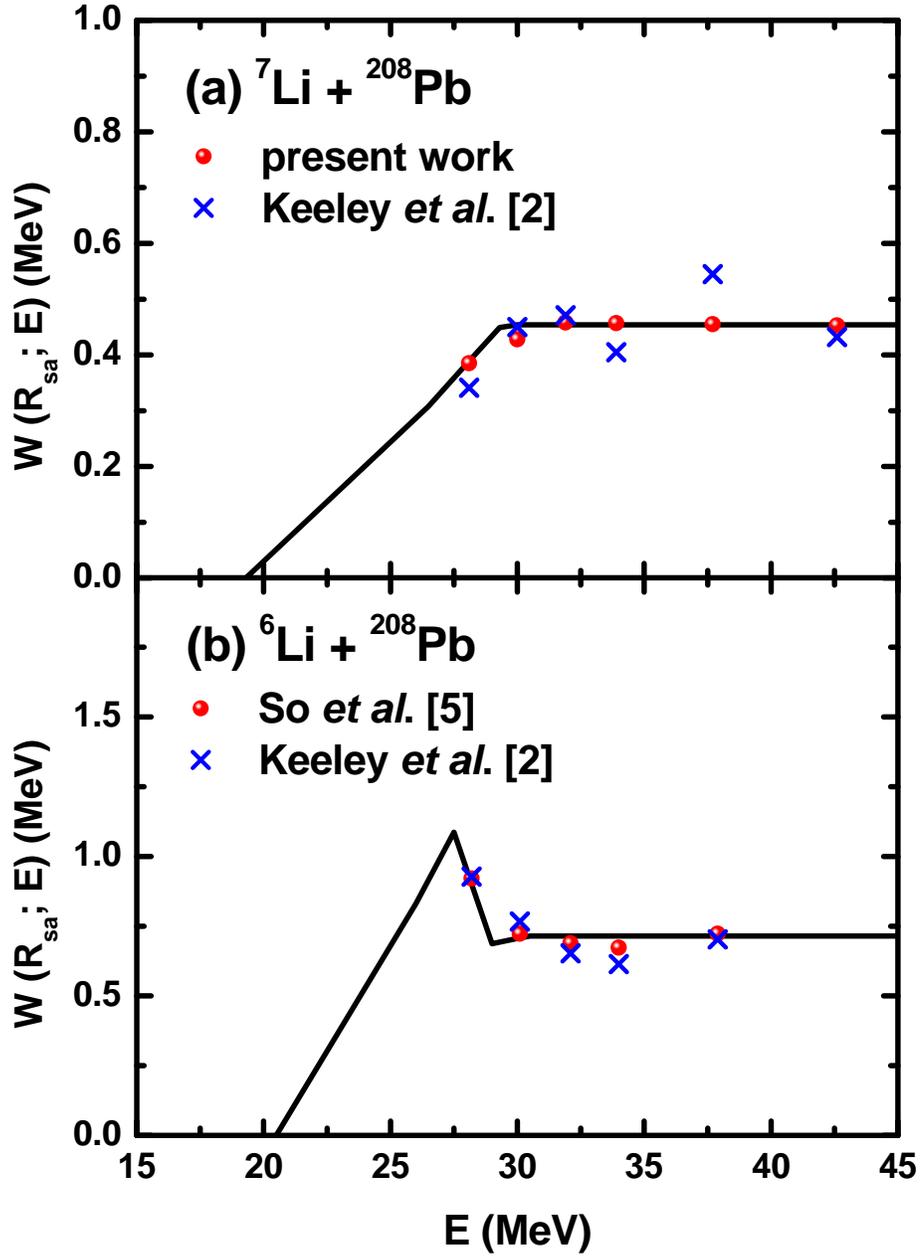}
\end{center}
\caption{\label{potential} (Color online) The values of the total
imaginary potential $W(r,E)$ at $r=R_{sa}=12.4$~fm deduced in the
present $\chi^{2}$ analyses and those obtained in
Ref.~\cite{kee1}. The values from Ref.~\cite{kee1} are multiplied by
factors 1.23 and 1.11 for $^{7}$Li+$^{208}$Pb and
$^{7}$Li+$^{208}$Pb system, respectively.}
\end{figure}
\end{document}